\documentclass[12pt,letterpaper]{article}

\usepackage{subcaption}

\usepackage{setspace} 
\usepackage{footmisc}
\usepackage{amsfonts}
\usepackage{amsmath}
\usepackage{theorem}

\usepackage[top=2.5cm, bottom=2.5cm, left=2.5cm, right=2.5cm]{geometry}

\usepackage{amsmath}
\usepackage{xcolor}

\newcommand\bovermat[2]{%
	\makebox[0pt][l]{$\smash{\overbrace{\phantom{%
					\begin{matrix}#2\end{matrix}}}^{\text{#1}}}$}#2}

\usepackage{pifont}

\usepackage{url}
\usepackage[T1]{fontenc}

\usepackage{setspace}
\usepackage{footmisc}

\usepackage[natbibapa]{apacite}
\usepackage{amssymb}
\setcitestyle{aysep={}} 

\usepackage{url}

\usepackage{mathtools}
\usepackage{amsmath}
\usepackage[latin2]{inputenc}
\usepackage{graphicx}
\usepackage{subcaption}
\usepackage{multirow}
\usepackage{esvect}
\usepackage{afterpage}

\usepackage{lipsum}
\usepackage{booktabs,array,dcolumn}

\usepackage{adjustbox}
\usepackage{graphicx}
\usepackage{pdflscape}

\usepackage{caption}
\usepackage[labelfont=bf]{caption}
\usepackage[figurename=FIGURE]{caption}
\usepackage[tablename=TABLE]{caption}

\let\ampersand\&
\renewcommand*\&{and}

\makeatletter
\def\@seccntformat#1{\@ifundefined{#1@cntformat}%
	{\csname the#1\endcsname\space}
	{\csname #1@cntformat\endcsname}}
\newcommand\section@cntformat{\thesection.\space}       
\newcommand\subsection@cntformat{\thesubsection.\space} 
\makeatother

\usepackage{atbegshi}
\AtBeginDocument{\AtBeginShipoutNext{\AtBeginShipoutDiscard}}

\begin{document}
	


\begin{titlepage}
\date{}

\begin{flushleft}
	
\title{\Large  \textbf{The iterative proportional fitting algorithm and the NM-method: solutions for two different sets of problems}}




\end{flushleft}

\begin{flushleft}
	
\singlespacing{\author{Anna NASZODI}}

\end{flushleft}
\thanks{Email: anna.naszodi@gmail.com.}\\

\maketitle
\thispagestyle{empty}

	
	\noindent 

\singlespacing{\textbf{Abstract:}  
	In this paper, we identify two different sets of problems.  
	The first covers the problems that the iterative proportional fitting (IPF) algorithm was developed to solve.  
	These concern completing a population table by using a sample.  
	The other set concerns constructing a counterfactual population table with the purpose of comparing two populations.  
	The IPF is commonly applied by social scientists to solve problems not only in the first set, but also in the second one.     
	We show that while it is legitimate to use the IPF for the first set of problems, it is not the right tool to address the problems of the second kind.  
	We promote an alternative of the IPF, the NM-method, for solving problems in the second set.  
	We provide both theoretical and empirical comparisons of these methods.}

\begin{flushleft}
\end{flushleft}

\end{titlepage}

\newpage

\section{Introduction}


The iterative proportional fitting (IPF) algorithm, or as it is also commonly referred to, the RAS algorithm,  
 is a mathematical scaling procedure which has been widely used by social scientists to standardize the marginal distribution of a contingency table (i.e., the so called seed table) to some fixed value, while
retaining a specific association between the row and the column variables. 
The retained association is the similarity of these variables captured by the odds-ratio.

Since there are a number of similarity indicators  
that are alternatives to the odds-ratio (e.g., the correlation coefficient, covariance, regression coefficient), 
the IPF table-transformation method also has a number of alternatives. 
These methods can be distinguished by the associations they  preserve.

In this paper, we \textit{compare the IPF with an alternative table-transformation method} that was recently proposed by \cite{NaszodiMendonca2021} (henceforth NM-method).\footnote{While the IPF is a built-in algorithm in SPSS, the NM is not yet.  
	However, there is no technical obstacle for the NM-method to gain popularity among non-SPSS users    
	since it is implemented in R, Excel and Visual Basic  
	(see: \url{http://dx.doi.org/10.17632/x2ry7bcm95.2}) and featured in the New York Times (see:  \url{https://www.nytimes.com/2023/02/13/opinion/marriage-assortative-mating.html}).   
	In addition, the NM-method is promoted in Wikipedia (see \url{https://en.wikipedia.org/wiki/NM-method}).}     
The transformed table obtained with the NM has the same preset marginal distributions as the transformed table obtained with the IPF. 
However, the NM-transformation is invariant to the ordinal indicator proposed by \cite{LiuLu2006} (henceforth, LL-indicator), rather than being invariant to the cardinal odds-ratio.

The \textit{main contribution of this paper is the clarification of the applicability of the odds-ratio-based IPF and the LL-indicator-based NM.}     
In particular, we show that {these methods are suitable to solve different problems:} 
the set of problems that the IPF can solve concerns \textit{completing a population table by using a sample}; 
the set of problems that the IPF cannot solve in general, while the NM can,  
concerns \textit{constructing a counterfactual population table} with the purpose of comparing two populations.

The limitation of the IPF's applicability  has been overlooked  as shown by its common use in social sciences for the second set of problems  (see e.g. \citealt{AlthamFerrie2007}; \citealt{BS2005}; \citealt{BreenSalazar2011}; \citealt{DupuyGalichon2014}; \citealt{HuQian2016}; \citealt{Leesch2022};  \citealt{Shen2021}).\footnote{This list of papers would be much longer if we also added those studies that rely either on the odds-ratio as a similarity indicator, or the odds-ratio based logistic regression.} 
The source of confusion is that the two types of applications are seemingly similar, although they are conceptually different.\footnote{Probably, due to the similarity of the two sets  of problems, the inventors of the IPF warned that  their  algorithm is ``not by itself useful for prediction''  (see \citealp{StephanDeming1940} p.444).} 
Also, the universal applicability of the IPF may be commonly thought to be justified by information theory.


In this paper, we introduce the well-known information theory-based justification of applying the IPF for the problems in the first set. 
Also, \textit{we contribute to the literature by developing the information theory-based justification of applying the NM for the problems in the second set}.

In addition, this paper  highlights some differences between the two sets of problems via two examples representing them.  
Most importantly, we argue that the stochastic variable in the problems of ``completing a population table'' has the role of \textit{selecting a random sample from a population}, 
while the stochastic variable in the problems of ``constructing a counterfactual population table'' \textit{selects the population table from the set of all possible population tables}.  


Another difference is the following.     
While the first set of problems concerns \textit{controlling for sample variation},   
the second one deals with \textit{constructing a counterfactual prediction}.\footnote{In some empirical applications, researchers have to deal with both kinds of problems. This is typically the case  if only a sample is observed of the population that the counterfactual population resembles. Even if these problems emerge jointly in some cases, these are two distinct problems.}

Finally, when constructing the counterfactual, the modeler can rely on all available information on the 
variables in the problem. 
When completing a population table, all the information to be used by the statistician are about the method of sampling on top of the information coded in the sample. 
For instance, if the modeler knows that a variable should exceed a certain threshold then he or she can explicitly take this restriction into account when constructing the counterfactual. By contrast, this restriction is taken into account only implicitly when completing the population table since the restriction is not violated by the sample.

As to the information theory-based justification of the IPF's applicability, we show  that once the Shanon entropy-based Kullback--Leibler (KL) directed divergence is decided to be used for quantifying the difference between the 
joint distributions represented by the seed table and by the transformed table in the first set of problems,  it is natural to apply the odds-ratio as a similarity indicator.  
Also, we argue that the LL-indicator, as well as the NM-method,  are consistent with a natural ranking of the possible joint distributions and a slightly modified version of the KL-divergence, the cumulative Kullback--Leibler (CKL) divergence.

Our paper defines the two distinct problems of  
``completing a population table by using a sample'' and  ``constructing a counterfactual population table''  not only in abstract terms but we also provide examples for them.  Our example for the first kind of problems is the simplest example that is typically solved by the maximum likelihood estimator. 
It concerns estimating the joint distribution of objects in a box by material and shape. 
For the estimation, we can use a random sample, as well as  the distribution by shape and the distribution by material of the objects in the box.

Our example for the second kind of problems is chosen from the literature on assortative mating.\footnote{However, it is worth noting that social sciences offer many similar examples.} 
It concerns using a counterfactual decomposition to compare the aggregate preferences for educational inter-marriages (or, the degree of sorting along the educational) in two consecutive generations, where the pair of educational distributions of marriageable men and marriageable women can be generation-specific.    

Hopefully, these examples help to guide social scientists at deciding whether to apply the IPF-algorithm together with the odds-ratio and the KL-divergence, or the NM-method together with the LL-indicator and the natural ranking to various table-transformation problems involving quantifying the strength of associations and comparing joint distributions.  
 
In the empirical part of this paper, we perform counterfactual decompositions with both methods. 
This application concerns documenting changes in the directly unobservable social cohesion among American young adults. Our observations are from four census waves, i.e., from 1980, 1990, 2000 and 2010. Accordingly, the observed individuals are typically from four different generations: in 1980 and 1990, most of the individuals observed are early boomers and late boomers, respectively. In 2000 and 2010, we observe mostly the members of the early generationX and the late generationX, respectively. 
We use their generation-specific  preferences for educational inter-marriages as an indicator for the strength of cohesion among different education groups. 

With this application, \textit{we illustrate that certain empirical findings are sensitive to the choice of the table-transformation method} (and it is also sensitive to the choice of  the similarity indicator and the choice of the indicator to compare distributions).  
While the IPF-based decomposition identifies the social cohesion among young adults in different education groups to have been \textit{monotonously weakened} between 1980 and 2010, the NM-method suggests that social cohesion displayed a \textit{hump-shaped pattern} over the same period. 
We use this disagreement of the IPF and the NM for method-selection:  
we present survey evidence about Americans' marital/ mating preferences that are consistent exclusively with the NM-implied hump-shaped pattern.  

The method-selection has practical consequences for evidence-informed policy-making.     
As an example, consider the possibility that the identified trends influence policy-makers' view about the effectiveness of certain policies at building social cohesion.        
Provided some scientific advisors identify the historical trends by using the IPF--odds-ratio--KL  triad, they find the generation of the late boomers  to be less cohesive  in 1990 relative to the early boomers in 1980 despite of the generous welfare policies that the late boomers could probably benefit from more than the early boomers.

Based on this trend ``evidenced'' by the combination of some data and some seemingly technical assumptions of the IPF--odds-ratio--KL triad, policy-makers are unlikely to view it possible to strengthen social cohesion in the future.   
Also, provided the IPF--odds-ratio--KL-user scientific advisors  ``document'' for the policy-makers that the late generationX has been similarly cohesive in 2010 as the early generationX was in 2000, the policy makers may acquire the misconception that not even the skyrocketing economic inequality during the Great Recession could damage social cohesion.     

However, the policy-makers' view on the future trends, as well as their view on the main risk to social cohesion, are likely to be different 
if they are offered reliable evidence for substantially stronger cohesion among the late boomers relative to the early boomers; and 
substantially weaker cohesion among the members of the late generationX relative to the early generationX.  
Actually, these trends are identified by any independent researcher, who applies the NM--LL--natural ranking triad rather than the IPF--odds-ratio--KL triad.

All in all, the significance of our survey-based method validation is due to the fact that the choice between the IPF and the NM is not innocuous:  
it makes a difference to what future paths are believed to be possible and also to what phenomena are believed to represent a major risk to a cohesive society.

As to the major source of risk, the literature finds inequality of various forms to positively associated with the degree of segmentation of the marriage market and to negatively associated with social cohesion. 
For instance, \cite{NaszodiMendonca2019} present evidence that the growing inequality of opportunity to find a job in the 2000s in the US  -- as well as in the European countries (France, Hungary and Portugal) they analyze  --  
coincided with the rise in inclination to choose a partner from  one's own education group.   
In particular, during the Great Recession, the employment prospects of people with low education level have worsened  relative to the highly educated individuals. At the same time, educational inter-marriages have typically became less desired in 2010 relative to 2000 in all the countries analyzed.  

\cite{NaszodiMendonca2019} identify the  employment gap--marital preferences association with time series data as well: they find Americans' stated marital preferences over spousal education monitored by  surveys to have had a more than 70\% correlation with the employment gap between highly educated and low educated workers between 1990 and 2017. 


For the methodological point we make, we exploit the strong negative correlation between social cohesion and economic inequality -- be it a causal relationship or not --  that makes the trend of inequality informative about the trend of cohesion.  
What regards the economic dimension of inequality, a consensus has been formed in the literature about its trend over the twentieth century.  
In particular, \cite{PikettySaez2003},  \cite{SaezZucman2016} document  \textit{income inequality} and \textit{wealth inequality}  to have a \textit{U-curve historical pattern} by using American tax records. {This pattern turned out to be robust to various modifications of the method and the data used: the stylized U-curve pattern could not be challenged.}     

This stylized pattern  is consistent with a hump-shaped pattern of cohesion under negative correlation.        
As we will see, the marriage market 
shows the strength of social cohesion 
to have displayed a hump-shaped pattern only if it is ``interrogated'' by the NM.    

Hopefully, this simple line of argument together with some theoretical considerations and the survey evidence we present in Subsection \ref{sec:Pew} 
will contribute to  
breaking the tradition of routinely applying the IPF in counterfactual decompositions to study assortative mating and similar phenomena (e.g. residential segregation, homophily in friendship formation, inter-generational mobility).\footnote{In this respect, our study  joins the seminal paper  by \cite{Mood2010}. Mood  aims at breaking the tradition of using  the odds-ratio-based logistic regression for comparing different groups.}   

The structure of this paper is the following. 
In Section \ref{sec:examp1}, we present an example for the kind of problems that the IPF is suitable to solve.  
In addition,  we motivate the application of the IPF in the related setting. 
In Section \ref{sec:cc}, we argue that  it is the NM-method  that provides the solution for another set of problems.     
Finally, Section \ref{sec:concl} concludes the paper.

\section{The problem of ``completing a population table''}\label{sec:examp1} 

Let us take an example to illustrate the problem of completing a population table by using a sample. 
There is a box full of different objects. The objects are tetrahedrons and cubes made out of either gold or silver.  
We know how many golden objects and how many argent objects are in the box.     
Moreover, we know how many tetrahedrons and how many cubes are in the box.    
We can observe a random \textit{sample} of these objects.
Our goal is to estimate the joint distribution of the \textit{population} of objects in the box by shape and material. 
 
We formalize this problem by using the following notations. 
We denote by $S$ and $B$ the tables of objects in the sample and in the box, respectively:\\
\begin{minipage}[p]{0.48\columnwidth}
  \centering
\begin{tabular}{llcc}  
	\multirow{2}{*}{\textit{S}} 	& & \multicolumn{2}{c}{Shape} \\ \cline{3-4}
	&	& Tetrahedron  & Cube \\ \cline{2-4} 
 \multirow{2}{*}{\rotatebox[origin=c]{90}{Material}} & 	Argent  & $n_{{\text{A}},{\text{T}}}$ & $n_{{\text{A}},{\text{C}}}$  \\ 
	& Gold & $n_{{\text{G}},{\text{T}}}$ & $n_{{\text{G}},{\text{C}}}$  \\  \cline{1-4}
\end{tabular}
\end{minipage}
\hfill
\begin{minipage}[p]{0.48\columnwidth}
\centering
\begin{tabular}{llcc|c}  
	\multirow{2}{*}{\textit{B}} 	 &	 & \multicolumn{3}{c}{Shape}   \\ \cline{3-5}
 &	& Tetrahedron  & Cube & Sum \\ \cline{1-5} 
\multirow{3}{*}{\rotatebox[origin=c]{90}{Material}} &	Argent  & $N_{{\text{A}},{\text{T}}}$ & $N_{{\text{A}},{\text{C}}}$  & $N_{{\text{A}},\cdot}$ \\ 
	&Gold & $N_{{\text{G}},{\text{T}}}$ & $N_{{\text{G}},{\text{C}}}$  & $N_{{\text{G}},\cdot}$ \\  \cline{2-5}
	&Sum & $N_{\cdot,{\text{T}}}$ & $N_{\cdot,{\text{C}}}$  & $N_{\cdot,\cdot}$ \\ 	 \cline{1-5}
\end{tabular}
\end{minipage}
\vspace{2mm}\\


We know the  number of objects of different types in $S$, i.e.,  
$n_{{\text{A}},{\text{T}}}$, $n_{{\text{A}},{\text{C}}}$,  $n_{{\text{G}},{\text{T}}}$ and  
$n_{{\text{G}},{\text{C}}}$, while we also know the 
row totals and column totals of $B$, i.e., the number of  
argent objects $N_{{\text{A}},\cdot}$, 
golden objects  $N_{{\text{G}},\cdot}$, 
tetrahedrons $N_{\cdot,{\text{T}}}$ and 
cubes  $N_{\cdot,{\text{C}}}$ in the box.  
We would like to estimate    
the number of  
argent tetrahedrons $N_{{\text{A}},{\text{T}}}$, 
argent cubes  $N_{{\text{A}},{\text{C}}}$,  
golden tetrahedrons $N_{{\text{G}},{\text{T}}}$ and 
golden cubes  $N_{{\text{G}},{\text{C}}}$  in the box.


A \textit{heuristically tempting approach} is to construct  table  $B$   
 by restricting  its  
odds-ratio (denoted by $\text{OR}^B$)  to be equal to the odds-ratio of the sample table $S$ (denoted by $\text{OR}^S$) 
  \begin{equation}\label{odds}
  \text{OR}^B= \frac{ N_{{\text{A}},{\text{T}}} N_{{\text{G}},{\text{C}}} } {N_{{\text{G}},{\text{T}}}  N_{{\text{A}},{\text{C}}} } = \frac{ n_{{\text{A}},{\text{T}}} n_{{\text{G}},{\text{C}}} } {n_{{\text{G}},{\text{T}}}  n_{{\text{A}},{\text{C}}} } = \text{OR}^S \;,
  \end{equation}
 while satisfying also the following four constraints given by the row-totals and column-totals:
 \small
 \begin{multline}\label{margin}{ 
N_{{\text{A}},\cdot}=  N_{{\text{A}},{\text{T}}}+ N_{{\text{A}},{\text{C}}}   \;,\;\; 
N_{{\text{G}},\cdot}=  N_{{\text{G}},{\text{T}}}+ N_{{\text{G}},{\text{C}}}   \;, \;\;
N_{\cdot, \text{T}}=  N_{{\text{A}},{\text{T}}}+ N_{{\text{G}},{\text{T}}}     \;,\;\; 
N_{\cdot, \text{C}}=  N_{{\text{A}},{\text{C}}}+ N_{{\text{G}},{\text{C}}}  \;.}
 \end{multline}
 \normalsize
 
If any three of the four restrictions in Eq.  (\ref{margin}) are fulfilled,  the fourth is automatically fulfilled as well, otherwise the 
different objects do not sum up to the total number of objects $N_{\cdot, \cdot}$. 
In other words, only three of the four restrictions are independent.  
Equation (\ref{odds}) complements the three independent  restrictions  with a fourth independent restriction. 
Since we have four independent restrictions and four unknowns, typically there is a unique solution to our problem.\footnote{An 
	exception is the special case  where at least one entry in the sample table is zero making it difficult to work with the odds-ratio.   
	This problem is only partially solved (see \citealp{Fienberg1970}).}

The next section justifies the heuristic approach. 
In addition, it introduces the IPF-algorithm and shows that this algorithm solves the problem of ``completing the population table''. Moreover, it shows that the IPF's  solution is identical to the solution obtained with the maximum likelihood estimator.

\subsection{Solving the problem of ``completing a population table''}\label{sec:OddIPF}

In this subsection we work with frequencies.   
Therefore, we norm tables $S$ and $B$ with the total number of objects in the sample ($n_{\cdot, \cdot}$) and the population  ($N_{\cdot, \cdot}$), respectively:
\begin{center}
	$Q=S/n_{\cdot, \cdot}=   \begin{bmatrix}
q_{A,T}    &  q_{A,C} \\
q_{G,T}   &  q_{G,C}
\end{bmatrix}   \;, \;\;\;\;\;\;\;\;\;\;\;\;\;\;\;\;\;\;\;\;\;$ 
$P= B/N_{\cdot, \cdot}=  \begin{bmatrix}
p_{A,T}    &  p_{A,C} \\
p_{G,T}   &  p_{G,C}
\end{bmatrix}   \;. $

\end{center}
%
%

In the problem of ``completing a population table'', we do not know $P$. 
However,  we know  its marginal  distributions given by  
$p_{\text{A},\cdot}= N_{{\text{A}},\cdot}/N_{\cdot, \cdot}$, 
$p_{\text{G},\cdot}= N_{{\text{G}},\cdot}/N_{\cdot, \cdot}$,   
$p_{\cdot,\text{T}}= N_{\cdot,{\text{T}}}/N_{\cdot, \cdot}$, 
$p_{\cdot, \text{C}}= N_{\cdot,{\text{C}}}/N_{\cdot, \cdot}$. 

We estimate $P$ by making it as close to the seed table $Q$ as possible. 
Let us measure their distance with the \textit{KL-directed divergence} and    
 perform the optimization over the set of distributions with preset marginal distributions.\footnote{So, even if we knew 
	that a variable to be estimated should exceed a certain threshold,  
we do not use this information since the only restriction we impose are on the marginal distributions.}    


    
The KL-divergence between the distributions $P$ and $Q$ is defined as
\begin{equation}\label{KL}
\text{DD}_{KL}(P||Q)=p_{A,T} \ln \left( \frac{p_{A,T}}{q_{A,T}} \right)+p_{A,C} \ln \left(\frac{p_{A,C}}{q_{A,C}}\right)+p_{G,T} \ln \left(\frac{p_{G,T}}{q_{G,T}}\right)+p_{G,C} \ln \left(\frac{p_{G,C}}{q_{G,C}}\right)  \;. 
\end{equation}

First, we note the following. 
If we knew $p_{A,T}$, we would also know   $p_{A,C}$, $p_{G,T}$, and $p_{G,C}$ since 
$p_{A,C}=p_{A,\cdot}-p_{A,T}$, 
$p_{G,T}=p_{\cdot,T}-p_{A,T}$  
and $p_{G,C}=p_{A,T}-p_{A,\cdot}+p_{\cdot,C}$.

By substituting these expressions into Equation (\ref{KL}), we get  
\begin{multline}\label{KL_x1}
\text{DD}_{KL}(P||Q)=p_{A,T} \ln\left( \frac{p_{A,T}}{q_{A,T}}\right)+
(p_{A,\cdot}-p_{A,T}) \ln\left(\frac{p_{A,\cdot}-p_{A,T}}{q_{A,C}}\right)+\\
(p_{\cdot,T}-p_{A,T}) \ln\left( \frac{p_{\cdot,T}-p_{A,T}}{q_{G,T}}\right)+
(p_{A,T}-p_{A,\cdot}+p_{\cdot,C}) \ln\left( \frac{p_{A,T}-p_{A,\cdot}+p_{\cdot,C}}{q_{G,C}} \right) \;. 
\end{multline}

We minimize $\text{DD}_{KL}(P||Q)$ by equating to zero its first derivative with respect to $p_{A,T}$: 
\begin{multline}\label{KL_x2}
0=\ln \left(\frac{p_{A,T}}{q_{A,T}}\right)  
-\ln\left( \frac{p_{A,\cdot}-p_{A,T}}{q_{A,C}}\right)  
-\ln\left(\frac{p_{\cdot,T}-p_{A,T}}{q_{G,T}}\right)
+\ln\left( \frac{p_{A,T}-p_{A,\cdot}+p_{\cdot,C}}{q_{G,C}}\right) \;. 
\end{multline}


By rearranging Eq. (\ref{KL_x2}), we obtain
\begin{equation}\label{KL_x3}  
\ln(q_{A,T})-\ln(q_{A,C}) -\ln(q_{G,T}) +\ln(q_{G,C}) = \ln(p_{A,T}) -\ln( p_{A,C})-\ln(p_{G,T}) +\ln(p_{G,C}) \;. 
\end{equation}

By taking the exponent of both sides of Eq. (\ref{KL_x3}), we get 
\begin{equation}\label{ORPQ}
\frac{q_{A,T} q_{G,C}}{q_{A,C}  q_{G,T}}= \frac{p_{A,T} p_{G,C}}{p_{A,C}  p_{G,T}}  \;, 
\end{equation}
where we have on the left-hand-side the odds-ratio (also called as the cross-product ratio) of the sample distribution table ($\text{OR}^Q$). 
And we have the odds-ratio of the population distribution table ($\text{OR}^P$)   on the right-hand-side.  
Trivially, the equality  $\text{OR}^S=\text{OR}^B$ also holds under Eq. (\ref{ORPQ}), because  $\text{OR}^Q=\text{OR}^S$ and $\text{OR}^P=\text{OR}^B$. 

This derivation proves that those methods that preserve the odds-ratio, minimize the KL directed divergence between the joint distributions represented by the 2-by-2 seed table and the 2-by-2 transformed table.   
As we will see next, the IPF algorithm is such.  


The IPF  is defined by the following two steps to be iterated until convergence (provided it converges).   
First, it factors the rows of the seed table $Q$ in order to match the row totals of $P$.  
The table obtained after  the first step (to be denoted by $Q^{'}$) may not have its column totals equal to the column totals of $P$. 
In this case, it is  necessary to perform a second step.         
As the second step, the IPF factors the columns of $Q^{'}$ to match the corresponding column totals of $P$. 
The table obtained  after  this  step (to be denoted by $Q^{''}$) may not have its row totals equal to the row  totals of $P$. 
In this case, repeating the first step is necessary with $Q=Q^{''}$. 
Alternatively, we stop the iteration. 

The table constructed by the IPF is the last value of $Q^{''}$. 
Let us denote it by $Q^{\text{IPF}}$. It has row totals that match the row totals of $P$, also  its column totals match the  column totals of $P$ by construction.  
More importantly, $Q^{\text{IPF}}$  has the same odds-ratio as  the seed table $Q$, because the odds-ratio is retained in each step of the algorithm:  
the ratio's numerator and denominator are multiplied by the same scalar in each step.\footnote{This property of the IPF is not new. It was already pointed out by \cite{Fienberg1970}.} 
These two properties of $Q^{\text{IPF}}$  prove that the IPF preserves the odds-ratio and thereby it minimizes the KL-divergence.

After visiting  the nexus among the odds-ratio, KL and IPF, let us see how these relate to the maximum likelihood estimator. 
The maximum likelihood estimator (MLE) solves exactly the type of problems that was illustrated by the example in the previous section.  
That is to estimate a population statistics, e.g., the population table, from the observations of a random sample. 
The MLE maximizes the conditional likelihood of observing sample $S$, where  $S$  is obtained by taking  $n_{\cdot, \cdot}$ number of independent random draws  from the population of $N_{\cdot, \cdot}$ number of objects.   
The MLE, similar to the IPF, minimizes the KL-divergence (see \citealp{Meyer1980}). 
Therefore, the IPF-transformed table is identical to the estimates obtained with the MLE.


Obviously, we could have presented some examples other than the one about the objects in a box.  
For instance, an example could have been the one concerning the joint educational distribution of wives and husbands in a sample of couples and the population of couples.\footnote{An other alternative example concerns the joint distribution of the four canonical nucleobases (Adenine, Guanine, Cytosine and Thymine) in the human DNA.} 
One can mechanically generate such an example by replacing the types of shapes and materials by the education levels of wives and husbands.
What is important to note however, is that our original example does not concern the (potentially stochastic) process determining the distribution of 
objects in the box. 
Similarly, the mechanically constructed example with couples does not concern  the process of couple-formation in a population.    
Therefore, we present, in the next section, an example with stochastic couple-formation.

\section{The problem of ``constructing a counterfactual table''}\label{sec:cc}

We illustrate the problem of ``constructing a counterfactual table'' with an example from  the literature on \textit{assortative mating} (see \citealp{LiuLu2006}, \citealp{NaszodiMendonca2021}).   
In this example, there are $N$ males and $N$ females in the population of a given generation. 
There are two educational categories, high ($H$) and low ($L$). 
The educational distributions are as follows. 
Among the $N$ men, $N_{H,\cdot}$ are high educated, while $N-N_{H,\cdot}$ are low educated.  
Among the $N$ women,   $N_{\cdot, H}$ are high educated. The other $N-N_{\cdot, H}$ are low educated.  
The educational distributions are assumed to be non-degenerate, i.e., 
$N>N_{H,\cdot}>0$ and $N>N_{\cdot, H}>0$.

Each marries someone from the opposite sex from his or her own generation.   
The joint educational distribution of the couples in the population of this generation is given by the 2-by-2 contingency table $K$:
\vspace{-8mm}
\begin{center}
	\begin{tabular}{llcc|c}  
	
	\multirow{2}{*}{\textit{K}} 		& 	 & \multicolumn{3}{c}{Wives}   
	\\ \cline{3-5}
  & 	& L  & H & Sum \\ \cline{1-5} 
\multirow{3}{*}{\rotatebox[origin=c]{90}{Husbands}}	& L  & $N_{L,L}$  & $N_{L,H}$  & $N_{L,\cdot}$ \\ 
	& H & $N_{H,L}$ & $N_{H,H}$   & $N_{H,\cdot}$ \\   \cline{2-5}
	& Sum & $N_{\cdot,L}$ & $N_{\cdot,H}$  & $N$ \\ 	 \cline{1-5}
\end{tabular}

\end{center}
where $N_{L,L}$ ($N_{H,H}$) denotes the number of homogamous couples where both spouses are low (high) educated. 
$N_{L,H}$ ($N_{H,L}$) stands for the number of heterogamous couples where the husband (wife) is low educated, while the wife (husband) is high educated.  

Now, suppose that we cannot observe the joint educational distribution of husbands and wives  in the population of a younger generation.  
However, we observe the educational distribution of marriageable men ($N^{\text{yg}}_{H,\cdot}$ and 
$N^{\text{yg}}_{L,\cdot}$)  and also the educational distribution of marriageable women  ($N^{\text{yg}}_{\cdot, H}$ and 
$N^{\text{yg}}_{\cdot, L}$) in this generation. 
These assumptions are realistic if members of the younger generation have already finalized their education but they have not started to form couples. 

We would like to know what the joint distribution of couples will be like in the younger generation if  wives' education will resemble their husbands' education in the younger  generation as much as these education levels resembled each other in the older generation.  
So, to solve the problem, we have to predict the following counterfactual table $K^{\text{yg}}$  
\begin{center}
	\begin{tabular}{llcc|c}  
				\multirow{2}{*}{$K^{\text{yg}}$} 		& 	 & \multicolumn{3}{c}{Wives}   	\\ \cline{3-5}
		 &	& L  & H & Sum \\ \cline{1-5} 
		\multirow{3}{*}{\rotatebox[origin=c]{90}{Husbands}}	& L  & $N^{\text{yg}}_{L,L}$  & $N^{\text{yg}}_{L,H}$  & $N^{\text{yg}}_{L,\cdot}$ \\ 
		& H & $N_{H,L}$ & $N^{\text{yg}}_{H,H}$   & $N^{\text{yg}}_{H,\cdot}$ \\   \cline{2-5}
		& Sum & $N^{\text{yg}}_{\cdot,L}$ & $N^{\text{yg}}_{\cdot,H}$  & $N^{\text{yg}}$ \\ 	 \cline{1-5}
	\end{tabular}
\end{center}
from its row totals, column totals and $K$. 

Obviously, the MLE is not applicable to solve this problem because this problem is different from estimating a population statistics from a sample. 
Maybe, the IPF, or an alternative to the IPF is applicable.   
We discuss the solution in the next section.

\subsection{Solving the problem of ``constructing a counterfactual table''}\label{sec:NotOddIPF}

Heuristically, to solve the problem of ``constructing a counterfactual table'',  we should choose  $K^{\text{yg}}$ as ``close'' to $K$ as possible. 
In  Subsection \ref{sec:OddIPF}, we quantified the closeness between two tables of the same size by the KL-divergence.  
Here, we use a different indicator, the LL-indicator that relies on a unique natural ranking of the tables. 

First, we introduce the concept of \textit{unique natural ranking} proposed by \cite{LiuLu2006}.  
It ranks  2-by-2 contingency tables with given row totals and column totals. 
Second, we define the degree-of-sorting measure  proposed by Liu and Lu. 
Then, we define the NM table-transformation method as the LL-indicator-invariant transformation.  
Finally, we make the point that the NM is consistent with the  \textit{cumulative residual entropy} (CRE) 
developed by \cite{Rao2004}.

It is well-known that a \textit{unique ranking} exists only over one-dimensional variables. 
For instance, there is no unique ranking over the 2-by-2 contingency tables with unknown row totals and column totals, because they  can be represented by a four-dimensional-variable.   
However, there is a unique natural ranking over the 2-by-2 contingency tables with given row totals and column totals, because these tables can be represented by a one-dimensional-variable.    
E.g.,  table 
$K$ can be represented by $N_{H,H}$ if   
the row totals and the column totals 
are known, because all the other three cells of $K$ are deterministic functions of  $N_{H,\cdot}$,  $N_{\cdot, H}$, $N_{L,\cdot}$,  $N_{\cdot, L}$ and $N_{H,H}$.      
The unique natural ranking over the one-dimensional values of the ${H,H}$ cells define the ranking over the 2-by-2 contingency tables with given row totals and column totals:   
that table ranks higher which has higher value in its ${H,H}$ cell.

We can normalize the rank by assigning the \textit{cumulative distribution} to each table provided the tables are uniformly distributed:  
e.g., if table $K$ ranks the 80th among all the 100 integer tables that have the same marginals as table $K$ itself then  
 we assign the value  0.8 to $K$.  

Alternatively, we can standardize the rank before we normalize it by assigning the value zero to those contingency tables that are equal to their (integer-valued) expected-value under independent row variable and column variable and given row total--column total pairs.  
Actually, this is how the ordinal measure proposed by Liu and Lu is constructed. 
In order to define it, we introduce some new notations.   

Let us denote by $R$ the expected value of the ${H,H}$ cell of those random non-negative integer tables, whose row sums and column sums are identical to the row sums and column sums of table 
 $K$ and whose row variable and column variable are independent.  
This expected value is calculated as $R={N_{H,\cdot}N_{\cdot,H}}/ {N}$. 
Similarly, we can calculate $R^{\text{yg}}={N^{\text{yg}}_{H,\cdot}N^{\text{yg}}_{\cdot,H}}/ {N^{\text{yg}}}$.

%
%

We assume that table $K$ is a random realization from the set $\{K_1 , \dotsc, K_k \}$ of all non-negative integer tables  with row sums  $N_{H,\cdot}, N_{L,\cdot}$      and column sums $N_{\cdot,H}, N_{\cdot,L}$. 
Similarly, table $K^{\text{yg}}$ is a random  realization from the set  $\{K^{\text{yg}}_1 , \dotsc, K^{\text{yg}}_l \}$ of all non-negative integer tables with row sums  $N^{\text{yg}}_{H,\cdot}, N^{\text{yg}}_{L,\cdot}$      and column sums $N^{\text{yg}}_{\cdot,H}, N^{\text{yg}}_{\cdot,L}$. 
So, in this example, a stochastic variable selects $K$ from an ordered set, while another stochastic variable selects $K^{\text{yg}}$ from another ordered set.

These stochastic variables are different from the stochastic variable in the example of objects of different materials and shapes. 
To recall, the latter variable selects a random sample from a population.

In the context of assortative mating, the association between the row variable (husbands' education level) and column variable (wives' education level) of each of the tables in the two sets is non-negative, because husbands and wives typically resemble each other in their educational trait. 
Therefore, it is reasonable to work with the proper subset $\{K^+_1 , \dotsc, K^+_{k+} \} \subset \{K_1 , \dotsc, K_k \}$ and 
the proper subset                                         $\{K^{\text{yg}+}_1 , \dotsc, K^{\text{yg}+}_{l+} \} \subset \{K^{\text{yg}}_1 , \dotsc, K^{\text{yg}}_l\}$,  where  
the value of the ${H,H}$ cells of each $K^+_i$, $i \in \{1 , \dotsc, k+ \}$ and each $K^{\text{yg}+}_j$, $j \in \{1 , \dotsc, l+ \}$   exceed certain thresholds.  
In particular, the values of the ${H,H}$ cells of tables in $\{K^+_1 , \dotsc, K^+_{k+} \}$ are assumed to be at least as high as  $\text{int}(R)$; while 
the values of the ${H,H}$ cells of tables in $\{K^{\text{yg}+}_1 , \dotsc, K^{\text{yg}+}_{l+} \}$ are assumed to be at least as high as  $\text{int}(R^{\text{yg}})$.


Moreover, irrespective of the sign of the association between the  row variable  and column variable  of the tables in the two sets, the values of the  ${H,H}$ cells of all the tables in $\{K^+_1 , \dotsc, K^+_{k+} \}$ and $\{K^{\text{yg}+}_1 , \dotsc, K^{\text{yg}+}_{l+} \}$  have their upper bounds, otherwise these  tables would not be non-negative.  
So, $\text{int}(R) \leq N_{H,H} \leq  \text{min}(N_{H,\cdot}, N_{\cdot,H})$ and $\text{int}(R^{\text{yg}}) \leq N^{\text{yg}}_{H,H} \leq  \text{min}(N^{\text{yg}}_{H,\cdot}, N^{\text{yg}}_{\cdot,H})$.


For table $K$ in the sub-set of 2-by-2 contingency tables with row totals $N_{H,\cdot}$,   $N_{L,\cdot}$,  column totals 
$N_{\cdot, H}$,  $N_{\cdot, L}$ and $\text{int}(R) \leq N_{H,H} \leq \text{min}(N_{H,\cdot}, N_{\cdot,H})$, the {LL-value} is defined as:   
\begin{equation}\label{LiuLusimpl} 
\text{LL}(K)=\frac{N_{H,H} - \text{int}(R)  }{\text{min}(N_{H,\cdot}, N_{\cdot,H} )-\text{int}(R) }. 
\end{equation} 
It is to be read as the actual value of cell ${H,H}$  minus its integer-valued expected value under independence over the maximum value of the same cell     
 minus its  minimum value.  

Accordingly, $\text{LL}(K)$ interprets as the percentile value of table $K$ among all the ranked tables in $\{K^+_1 , \dotsc, K^+_{k+} \}$.  
Similarly, $\text{LL}(K^{\text{yg}})=({N^{\text{yg}}_{H,H} - \text{int}(R^{\text{yg}})  })/({\text{min}(N^{\text{yg}}_{H,\cdot}, N^{\text{yg}}_{\cdot,H} )-\text{int}(R^{\text{yg}}) })$ interprets as the percentile value of table $K^{\text{yg}}$ among all the ranked tables in $\{K^{\text{yg}+}_1 , \dotsc, K^{\text{yg}+}_{l+} \}$.  


According to the natural-ranking-based definition of closeness, $K^{\text{yg}}$ is closest to $K$, iff 
\begin{equation}\label{eq:LLEq} 
\text{LL}(K^{\text{yg}})=\text{LL}(K). 
\end{equation} 


 
Similarly to the KL-based (or odds-ratio-based) concept of closeness, the natural ranking-based (or LL-based) concept of closeness  also defines a unique method for transforming any 2-by-2 contingency table $K$ to a 2-by-2 contingency table  with preset row totals and column totals. 
The reason is the same: the restrictions on the row totals and column totals represent three independent restrictions, while  Eq. (\ref{eq:LLEq}) represents the fourth independent restriction.  These four restrictions determine the values of each of the four cells of the 2-by-2 transformed contingency table.  

If we define closeness with the LL-value (rather than with the KL-divergence), then the table-transformation method is the one proposed by \cite{NaszodiMendonca2021} (rather than the IPF), since it preserves the LL-value of the table to be transformed by definition. 

Next, let us visit a theoretical argument in favor of the  application of the   NM for constructing counterfactuals: 
we show that similarly to the IPF--odds-ratio--KL triad, the application of the NM can also be backed by information theory. 
In particular, we  introduce  how the natural-ranking-based (or LL-based) definition of closeness relates to an entropy-measure. 


Let us denote by  $F^{\text{og}}$ the cumulative distribution function of $\text{LL}(K)$: it is the probability of observing a table $K^{\text{r}} \in \{K^+_1 , \dotsc, K^+_{k+} \}$  where $\text{LL}(K^{\text{r}})\leq \text{LL}(K)$. Distribution $F^{og}$ may not necessarily be uniform: some population tables in $\{K^+_1 , \dotsc, K^+_{k+} \}$ can be assumed to be more likely to realize than some others. 
Further, let us denote by  $F^{\text{yg}}$ the cumulative distribution function of $\text{LL}(K^{\text{yg}})$: it is the probability of observing a table $K^{\text{s}} \in \{K^{\text{yg}+}_1 , \dotsc, K^{\text{yg}+}_{l+} \}$  with $\text{LL}(K^{\text{s}})\leq \text{LL}(K^{\text{yg}})$. 
  
\cite{BaratpourRad2012} introduce the CKL as the distance between distributions that builds on the CRE. In particular, if $X$ and $Y$ are two non-negative  random variables with cumulative distribution functions $F$ and $G$, and probability density functions $f$ and $g$ then the CKL defines the distance between the distributions $F$ and $G$.
\cite{BaratpourRad2012} show in a lemma that  $\text{CKL}(F:G) \geq 0$ and equality holds if and only if the \textit{cumulative distributions} are equal $F=G$ almost everywhere. 

By applying this lemma to our problem, 
$\text{CKL}(F^{\text{yg}}:F^{\text{og}}) \geq 0$ and equality holds iff the \textit{cumulative distributions} are equal $F^{\text{yg}}=F^{\text{og}}$ almost everywhere. The equality of the cumulative distributions holds under  $\text{LL}(K^{\text{yg}})=\text{LL}(K)$. Thereby, we have proven that any table transformation method that preserves the LL-indicator, minimizes the CKL. The  NM preserves it by definition, therefore it  minimizes $\text{CKL}(F^{\text{yg}}:F^{\text{og}})$.  

It is worth to note that there is a similar lemma for the KL-divergence.   
In particular, $\text{DD}_{\text{KL}}(F||G)\geq 0$ as well. 
The equality holds iff the \textit{probability density functions}, i.e.,  $f$ and $g$ are equal almost everywhere. 

The advantage of the CKL relative to the KL is that the former is applicable even if we do not know the probability density functions (either because they do not exist, or because  our knowledge is limited in this respect), while we know the cumulative density functions. 
This is often the case in practice. For instance, a weather forecaster may have only limited knowledge about the probabilities of various temperatures tomorrow, while he or she can  rank the probabilities of having the temperature under different thresholds. Similarly, it is much easier for social scientists to rank different counterfactuals than to assign probabilities to them. In addition, ranking the counterfactuals is sufficient for comparing different populations if the method for constructing  counterfactuals is the NM.

In the next section, we compare the empirical performances of the  IPF and the NM at constructing counterfactuals.  
As we will see, only the outcome of the  NM-based counterfactual decompositions are in accord with some survey evidence.

\subsection{Empirical application of the IPF and the NM}\label{sec:diff_emp}

In our empirical application, we perform counterfactual decompositions both with the IPF and the NM.   
This application concerns studying the changes in aggregate revealed marital/ mating preferences over inter-educational relationships.\footnote{In this respect, we deviate from the paper by  \cite{NaszodiMendonca2021} that illustrates the application of the NM-method by studying the aggregate preferences for educationally homogamous relationships.}  
We work with US census data about the joint educational distributions of American heterosexual couples  in 1980, 1990, 2000 and 2010.   

The couples observed are either officially married, or unmarried cohabiting romantic partners.   
Following \cite{NaszodiMendonca2021}, we restrict our analysis to young couples defined by the age of the male partner. 
In particular, all the husbands/ male partners in our data are between 30 and 34 years old when being observed. 
This choice guarantees that almost all the husbands/ male partners and almost all the wives/ female partners in our data 
have already attained their final education level when being observed.        
Also, our choice of the age group analyzed can be motivated by the fact that we do not have overlapping observations under our definition of young couples, i.e., not even the most steady relationships enter our decennial data more than once.

We distinguish between three education levels: ``Low'' (corresponding to no high school degree), ``Medium'' (corresponding to high school degree without tertiary level diploma), and ``High'' (corresponding to tertiary level diploma). Accordingly, the educationally heterogamous couples (of husbands/ male partners -- wives/ female partners) are either of the following six types: Low--Medium, Low--High, Medium--High, Medium--Low, High--Low, and High--Medium. 

As part of our empirical analysis, we perform the following three steps.  
First, we calculate the share of educationally heterogamous couples  among all couples in each of the four census years. 
Then, we decompose the changes in the share of heterogamous couples between 1980 and 1990, as well as between 1990 and 2000, 2000 and 2010 into 
(i) the ceteris paribus effects of changing marginal distributions from one generation to the next generation,    
(ii) the ceteris paribus effects of changing aggregate preferences from one generation to the next generation,\footnote{While we refer to these effects as the effects of changing aggregate preferences, these could also be called as the effects of changing marital social norms, or social barriers to marry out of ones' group, or social gap between different groups, or  degree of segmentation of the marriage market, or  degree of sorting. Irrespective of the terminology used, these effects are empirically equivalent to each other provided they are all identified from the changes in the  matching outcome by controlling for changes in the structural availability of marriageable men and women with various traits.} 
 and   
(iii) the joint effects of changing preferences and marginal distributions. 

In one of the specifications, we construct the counterfactuals for the decompositions with the IPF, whereas we construct them with the NM in the alternative specification.   
For more details about how we perform the counterfactual decompositions see either the Appendix, or \cite{NaszodiMendonca2021}. 

Finally, we use 1980 as a reference year and calculate what would have been the share of heterogamous couples like in all the other years  
if only the preferences had changed within each decade.  
 

\begin{figure}
	\begin{center}
		\caption{Share of heterogamous American couples between 1980 and 2010}
		\includegraphics[width=.95\linewidth]{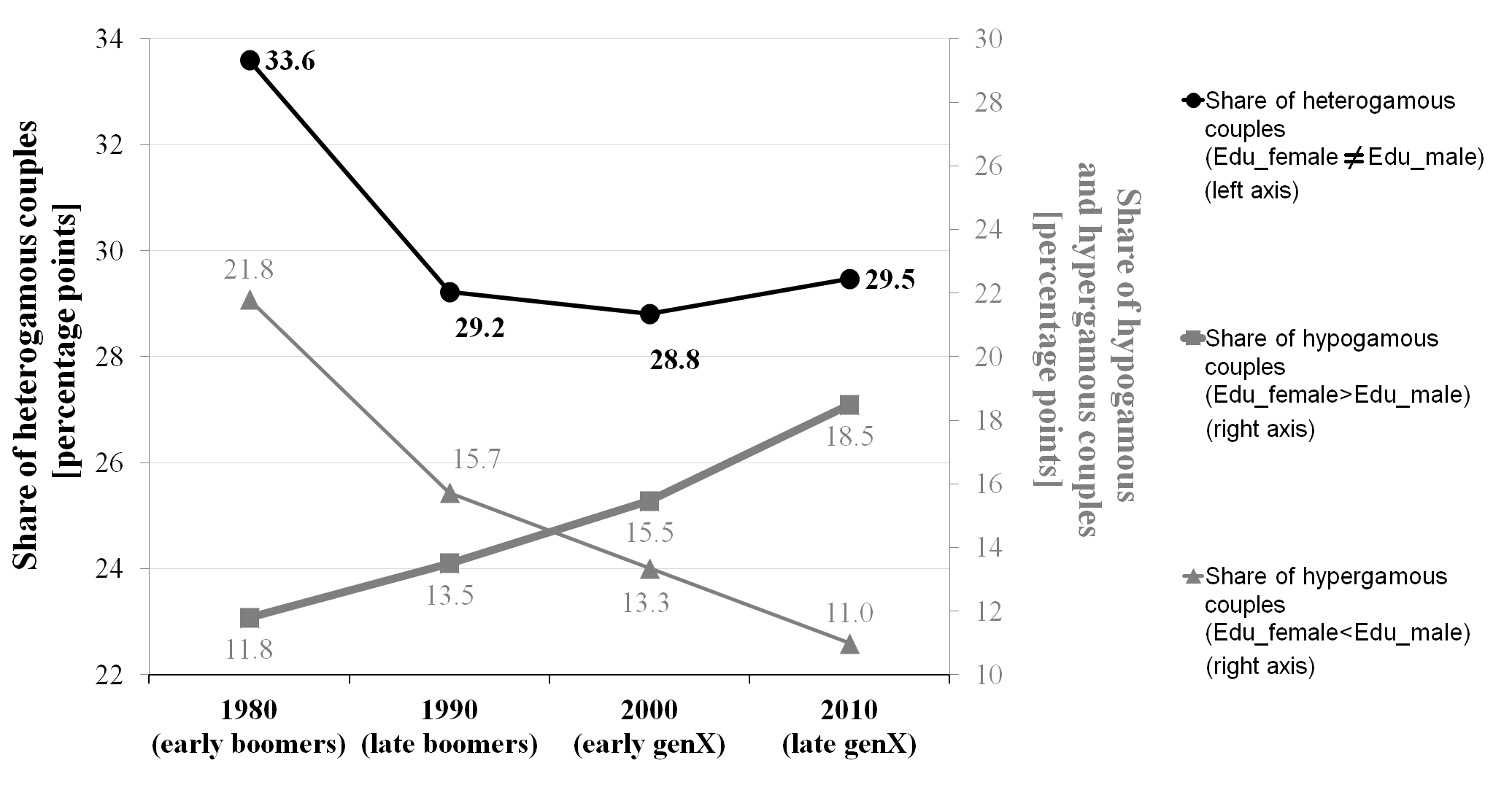}
		
		\label{fig:US_heterog_hist}
	\end{center}
	\textit{Source}: author's calculations based on US \textit{census data} from IPUMS about the education level of married couples and cohabiting couples in 1980, 1990, 2000 and 2010.\\
	\textit{Notes}: The husbands/male partners were [30,34] years old when observed. Those observed in 1980 and 1990 belong to the generation of the early boomers and late boomers, respectively; in 2000 and 2010, the observed husbands/male partners were members of the early generationX and the late generationX, respectively. The educational attainment can take three different values: ``Low'': no high school degree, ``Medium'': high school degree without tertiary level diploma, and ``High'': tertiary level diploma. Accordingly, the educationally hypogamous couples (of husbands/male partners--wives/female partners) are either Low--Medium, or Low--High, or Medium--High types of couples. While the educationally hypergamous couples are either Medium--Low type, or High--Low type, or High--Medium type. 
	
\end{figure}

\begin{figure}
	\begin{center}
		\caption{Some gender-ratios potentially effecting the opportunities of heterogamous couple-formations}
		\includegraphics[width=.95\linewidth]{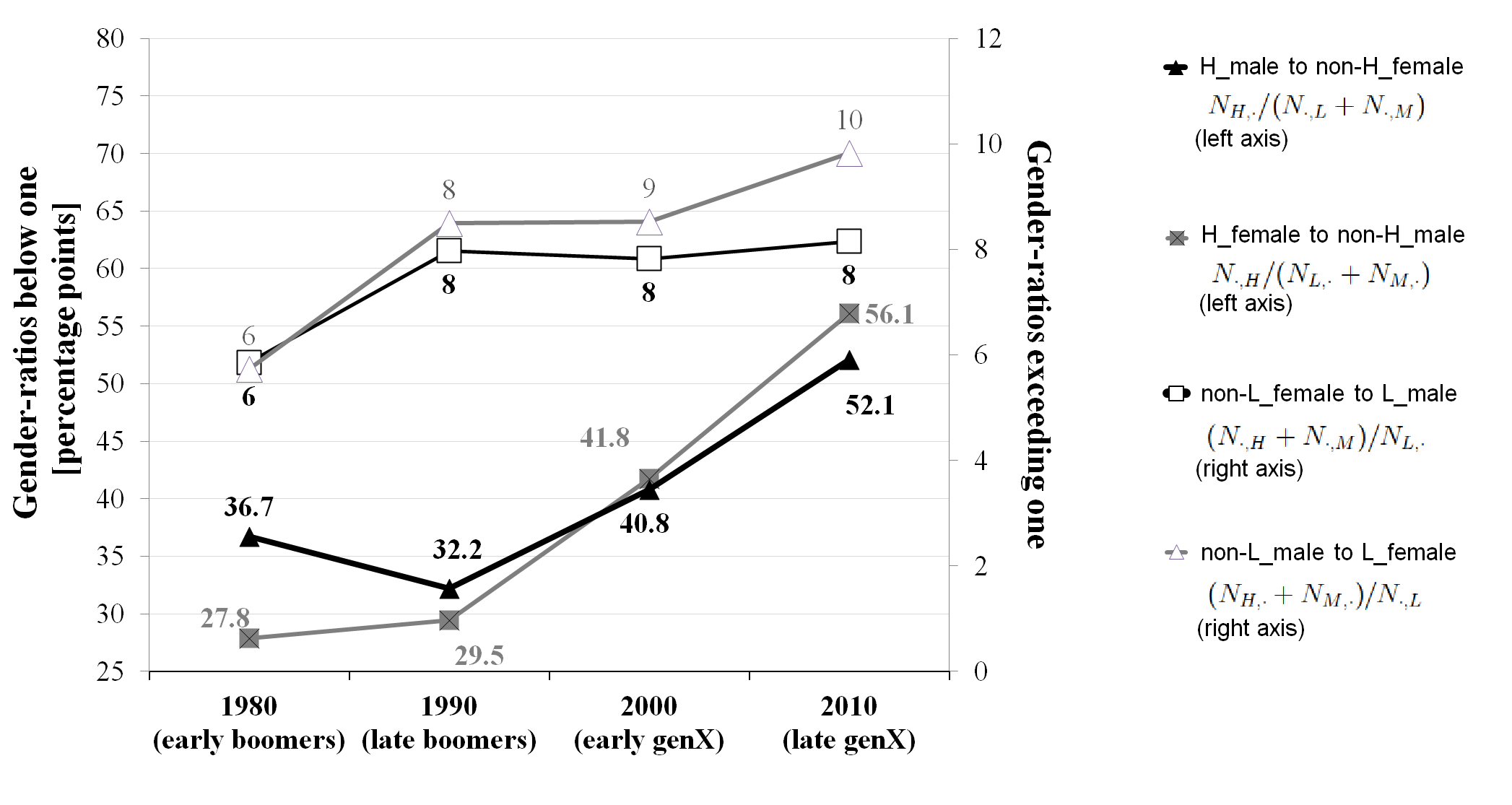}
		
		\label{fig:US_genderratio_hist}
	\end{center}
	\textit{Notes}: same as under Figure \ref{fig:US_heterog_hist}.  
	
	
\end{figure}

Figure \ref{fig:US_heterog_hist} presents the time series of the share of heterogamous couples together with its break-down to hypergamous couples (where the education level of the male partners exceeds that of the female partners) and hypogamous couples (where the female partners are more educated).

Apparently,  the share of hypogamous couples  and the  share of hypergamous couples displayed the opposite trends over the  decades studied.   
Whereas their sum  decreased substantially between 1980 and 2000 (by more than 4 percentage points, from 33.6\% to 28.8\%), after which it increased slightly 
 between 2000 and 2010 (by 0.7 percentage points). So, the prevalence of educational heterogamy displayed a U-shaped pattern over the three decades analyzed.  
 This pattern itself is limitedly indicative about the changes in the factor of our interest, i.e., the preferences for educational heterogamy.   
 The reason is that the historical trend of the prevalence of inter-educational relationships depends not only on the desires, but also on the opportunities and their interaction (see \citealp{Kalmijn1998}).

 Figure \ref{fig:US_genderratio_hist} shows the evolution of four education level-specific gender-ratios. 
 The dynamics of one of these ratios may be especially relevant: 
  the decline in  $N_{H,\cdot}/(N_{\cdot,L}+N_{\cdot,M})$ 
   over the 1980s  may have remarkably limited the opportunities of women with no tertiary education diploma to marry up. Whereas its positive trend after 1990 may have enhanced their opportunities.  
    
 The other three gender-ratios $N_{\cdot,H}/(N_{L,\cdot}+N_{M,\cdot})$, $(N_{\cdot,H}+N_{\cdot,M})/N_{L,\cdot}$ and $(N_{H,\cdot}+N_{M,\cdot})/N_{\cdot,L}$ displayed a monotonous  positive trend over all the three decades analyzed. So, provided these trends exerted any effect, their effect was positive on the opportunities of heterogamous couple-formations.

 

\begin{figure}
	\begin{center}
		\caption{Decomposition of changing prevalence of {educational heterogamy}}
		\centering
		\includegraphics[width=.95\linewidth]{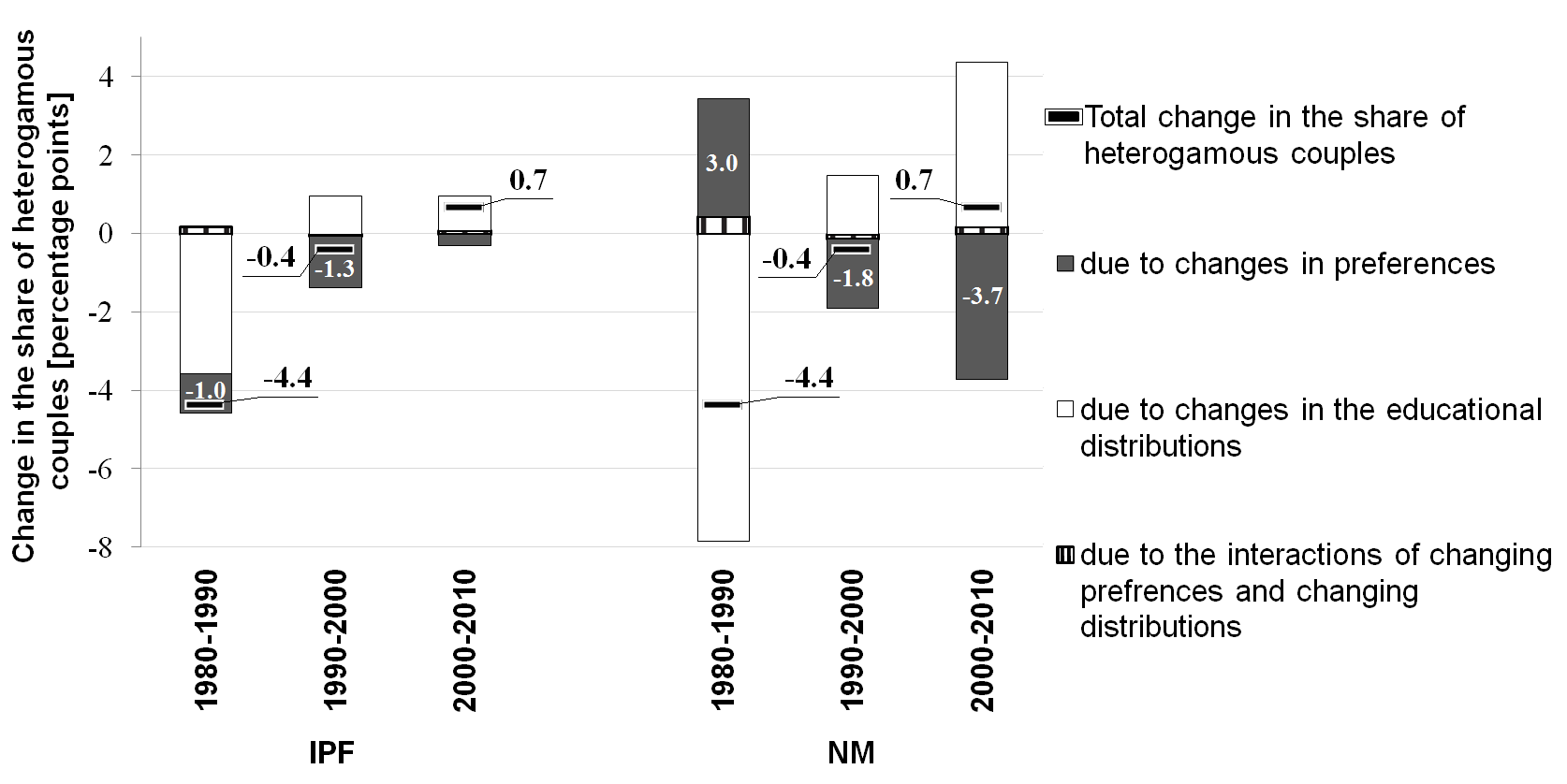}
		
		\label{fig:US_decomp_heterog_IPF_NM}
	\end{center}
	\textit{Notes}: same as under Figure \ref{fig:US_heterog_hist}.
	Changes in the prevalence of heterogamy are  decomposed by using the so called {additive decomposition scheme with interaction effects}  (see Eq. \ref{eq:Bdecom2}), while the counterfactual tables are constructed either with the IPF, or the NM.	 
	
\end{figure}

Next, let us turn to the results of the decompositions.      
Figure \ref{fig:US_decomp_heterog_IPF_NM} presents the outcome of the decompositions performed with the IPF and the NM.  
Our focus is on the components representing the effects of changing aggregate preferences (see the dark bars of Fig. \ref{fig:US_decomp_heterog_IPF_NM}), since these components proxy the changes in the share of social cohesion among different education groups.    
These components are not robust to the method used for constructing the counterfactuals. 
In particular, the components belonging to the 1980s and the first decade after the turn of the Millennium are especially sensitive.

\begin{figure}
		\caption{Share of educationally heterogamous couples under the counterfactual of no change in the educational distributions between the census years} 
	
	\begin{subfigure}{0.48\textwidth}
		\includegraphics[width=\linewidth]{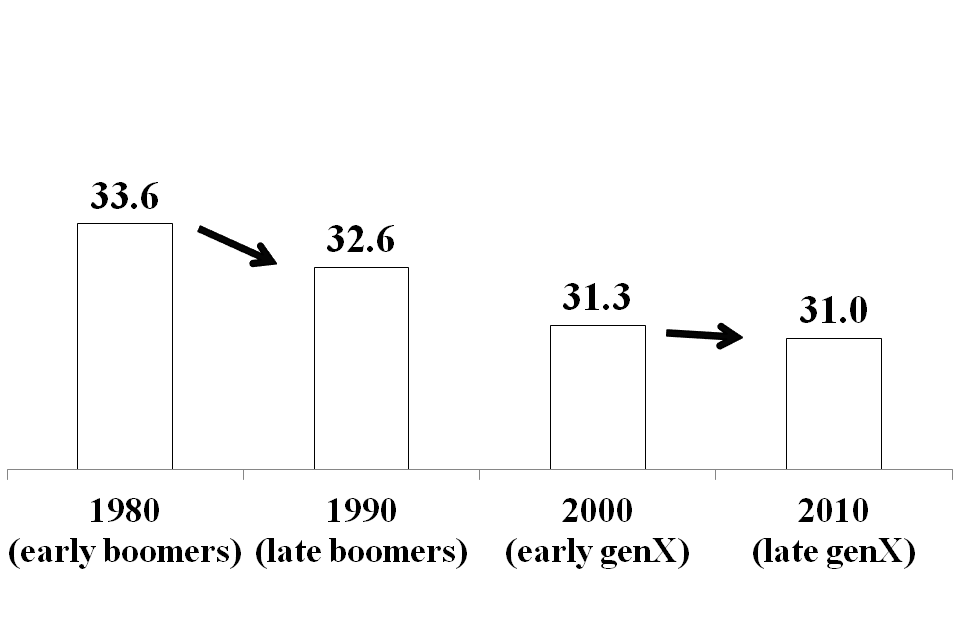}
		\caption{Counterfactuals constructed with IPF}
		\label{fig:IPF}
	\end{subfigure}%
	\hspace*{\fill}   
	\begin{subfigure}{0.48\textwidth}
		\includegraphics[width=\linewidth]{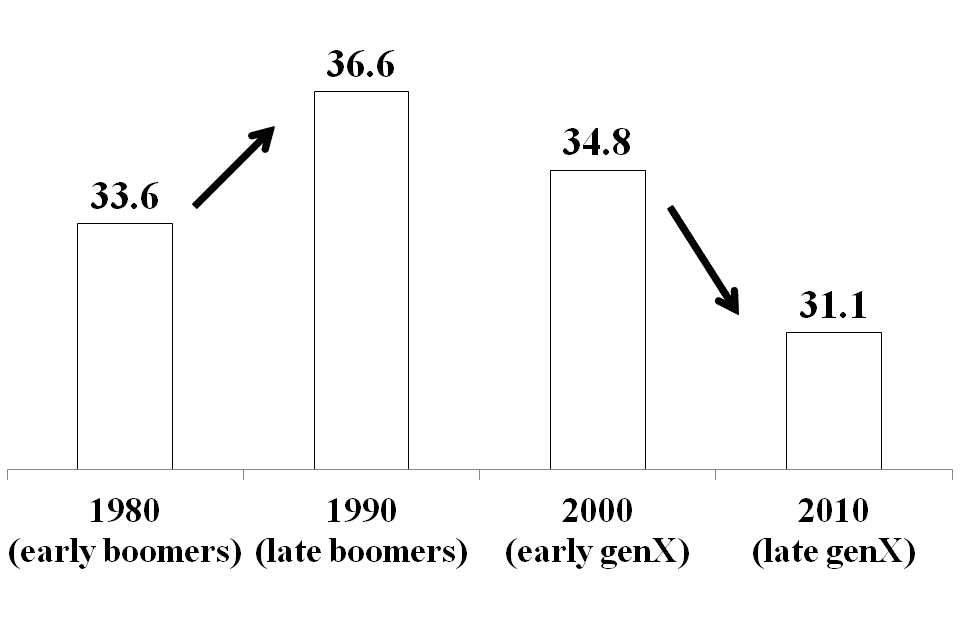}
		\caption{Counterfactuals constructed with NM}
		\label{fig:NM}
	\end{subfigure}\\%
	\textit{Notes}: same as under Figure \ref{fig:US_decomp_heterog_IPF_NM}. We assign the historical value to the reference year of 1980; to all the other years,  we assign  their counterfactual values (i.e., the historical value in 1980 plus the cumulative contributions of the changes in aggregate marital preferences).   
	\label{fig:SHetC_19802010}
\end{figure}

We use the components to construct the time series of the share of heterogamous couples under the counterfactual of unchanged opportunities within each decade. 
Since the components themselves are sensitive to the choice between the IPF and the NM, the counterfactual share of heterogamous couples is also sensitive to the choice of the method  (see Figure \ref{fig:SHetC_19802010}).  


By considering  the effect of changing aggregate preferences to be an indicator of social cohesion, we interpret the outcome of the decomposition as follows.  
The IPF-based decomposition suggests that the generation of late boomers was substantially \textit{less} cohesive along the educational dimension in 1990 than the generation of the early boomers was in 1980. The NM-based decomposition suggests the opposite. 
Also, the IPF-based decomposition suggests that the late generationX was almost as cohesive in 2010  as the  early generationX was in 2000. 
By contrast, the NM-based decomposition documents that the late generationX was much less cohesive relative to the early generationX.

\subsection{Validating the application of NM with survey evidence}\label{sec:Pew}

We validate the approach of constructing counterfactuals with the NM by using survey evidence on changes in Americans' marital preferences.  
In particular, we use the Pew Research Center's survey called Changing American Family Survey from 2010. 

Its question Q.23F1 (Q.24F2) asked from female (/male) participants was the following. ``People have different ideas about what makes a man (/woman) a good husband (/wife) or partner. For each of the qualities that I read, please tell me if you feel it is very important for a good husband (/wife) or partner to have, somewhat important, not too important, or not at all important. First, he (/she) is well educated. Is this very important for a good husband (/wife) or partner to have, somewhat important, not too important, or not at all important?''  

We estimate the population-shares of those men and women in each of the four generations analyzed, who viewed spousal education \textit{not to be important at all}. 
We refer to this share as the share of very inclusive individuals.   

For the estimations, we apply the approximation proposed by \cite{AgrestiCoull1998}. 
Following them, we assume that the distribution of the number of survey responses ``not at all important'' (denoted by $x$) out of $n$ number of total responses is binomial with the parameter $PS$. 
While $q = x/n$ is the sample-share of very inclusive respondents, $PS$ denotes their population-share. 
\cite{AgrestiCoull1998} propose to estimate the latter as
\begin{equation}\label{eq:AC} 
\widehat{PS} = (x + z^2/2)/(n + z^2) \;, 
\end{equation} 
where $z = \Phi^{-1}(1-\alpha/2)$ is the quantile of a standard normal distribution with $\alpha = 5\%$.  

Also, they propose to approximate the population-share's symmetric confidence interval by
\begin{equation}\label{eq:ACconfint}
\widehat{PS} \pm z \sqrt{\hat{p}(1-\hat{p})/(n+z^2) } \;. 
\end{equation} 

\begin{figure}
	\caption{Generation-specific views of the opposite sex on the (un)importance of spousal education in the US in 2010.} 
	
	\begin{subfigure}{0.48\textwidth}
		\includegraphics[width=\linewidth]{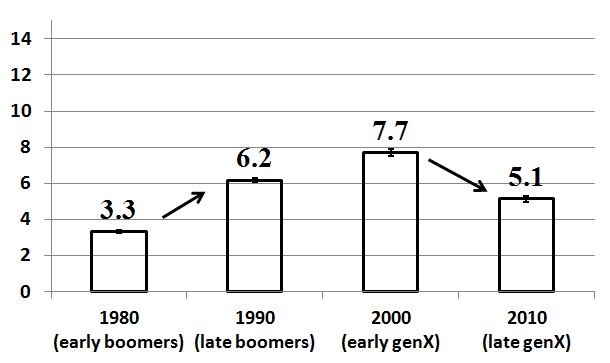}
		\caption{Proportion of \textit{women} saying it is a ``not at all important quality'' of a good husband/partner to be well-educated}
		\label{fig:Pew_female}
	\end{subfigure}%
	\hspace*{\fill}   
	\begin{subfigure}{0.48\textwidth}
		\includegraphics[width=\linewidth]{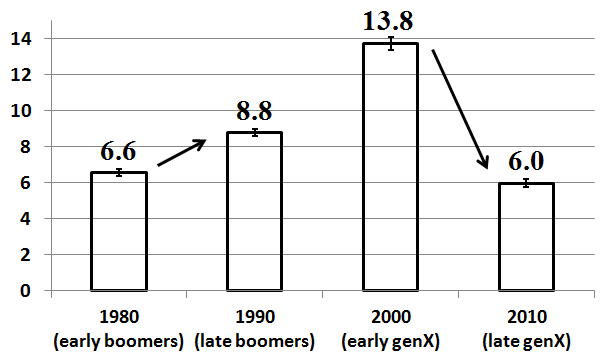}
		\caption{Proportion of \textit{men} saying it is a ``not at all important quality'' of a good wife/partner to be well-educated}
		\label{fig:Pew_male}
	\end{subfigure}\\ %
	
	\textit{Source}:  author's calculations based on the answers to the \textit{survey} questions number 23F1 and number 24F2 in the 
	Changing American Family survey.\\
	\textit{Notes}:  On the horizontal axis, we see the census year in which the survey respondents were 30--34 years old. 
	The chart is based on the answers of 84 female and 56 male respondents from age group 60-64 (representing early boomers),  92 female and 75 male respondents from age group 50-54 (representing  late boomers),  60 female and  61 male respondents from age group 40-44 (representing early genX),  53 female and 45 male respondents from age group 30-34 (representing late genX). 
	The population-shares, as well as the 95\% symmetric confidence intervals, are estimated with the approximation proposed by \cite{AgrestiCoull1998} (see Equations \ref{eq:AC}, \ref{eq:ACconfint}). 
	\label{fig:Pew_disagree}
\end{figure}

The results of our survey data  analysis are presented by Figure \ref{fig:Pew_disagree}. 
Apparently, the population-share of very inclusive individuals among the late boomers was  higher relative to  the early boomers. 
This trend is qualitatively the same for females and males (see Figure \ref{fig:Pew_disagree} a, b). 
Moreover, the inter-generational difference is significant since the confidence intervals are disjunct (see the short line segments at the top of the bars of Fig.  \ref{fig:Pew_disagree}). 
As to the generationX, we find the opposite trend: the population-shares of very inclusive men and women  were significantly lower among the members of the late generationX than among the members of the early generationX. 





These findings also translate into a hump-shaped pattern of social cohesion.     
Therefore, the survey evidence corroborates  the application of the NM for constructing counterfactuals.\footnote{While the hump-shaped curve is robust to being identified from the survey data or  the census data ``interrogated'' by the NM, its turning point is somewhat sensitive to the choice of the data.} 

\section{CONCLUSION}\label{sec:concl}

In the theoretical part of this paper, we highlighted the difference between the two sets of problems that the IPF-algorithm  and the NM-method were original developed for. 
Moreover, we offered some guidance for social scientists at deciding whether to apply the IPF, or the NM to various table-transformation problems. 
In particular, we presented some theoretical considerations that are in favor of applying the IPF for completing a population table by using a random sample from the population. 
Also,  we presented some theoretical considerations  that are in favor of applying the NM for constructing counterfactuals. 


In the empirical part of this paper, we studied changes in social cohesion in the US via analyzing Americans' generation-specific aggregate preferences for inter-educational marriages between 1980 and 2010. We used this application to compare the performances of the competing methods at conducting counterfactual decompositions.  
The two methods turned out to disagree about the historical trend of social cohesion.    

Finally, we conducted an empirical method-selection by using survey evidence about American's declared marital preferences. 
Both the survey evidence and the NM-based counterfactual decomposition suggest that the American late boomers were typically more permissive towards inter-educational couple-formation relative to the early boomers, while the members of the late generationX were typically less permissive towards inter-educational relationships than the members of the early  generationX.
 
These findings are also consistent with the U-shaped pattern of income inequality and wealth inequality  documented by \cite{PikettySaez2003},  \cite{SaezZucman2016}.
However, none of these empirical evidence corroborate the application of the IPF for constructing counterfactuals.  
What can social scientists do about it?\footnote{A closely related question is raised in the title of the paper by \cite{Mood2010}.}   
The answer is simple: we can use the NM instead of the IPF in applications aiming at comparing different populations via counterfactual decompositions.

\newpage

{\Large \textbf{Appendix}}

This appendix offers a detailed explanation on how the empirical decompositions are performed. 
First, we define the generalized LL-indicator for contingency tables larger than 2-by-2. 
Second, we present the  NM-method. In addition, we illustrate its application and the application of the IPF with a numerical example.   
Finally, we present the decomposition scheme used. \\

\textbf{The generalized LL-indicator}

The first thing to note is that the LL-value, as well as  the odds-ratio,      
are defined for 2-by-2 contingency tables by Equations  (\ref{LiuLusimpl}) and (\ref{odds}). 
However, in the empirical part of the paper we work with a multinomial assorted trait variable as the education level can take 3 different values.  
 Here, we  {relax the assumption that the assorted trait is dichotomous}.

In the \textit{multinomial case}, the one-dimensional assorted trait distribution can even be gender-specific. 
For instance, it is possible that the market distinguishes between $m\geq2$ different education levels of women and  $n \geq 2$  different education levels of men where $n$ may not be equal to $m$.   
Let us denote by $Z_t$  the contingency table (of size  $n \times m$) representing the aggregate market equilibrium at time $t$. 
(Here, we deviate from the notation used in Section \ref{sec:cc} by introducing the time index.) 

If both the male-specific assorted trait variable and the female-specific assorted trait variable are {one-dimensional, ordered, categorical, multinomial variables}  
then the aggregate marital preferences at time $t$ can be characterized  by the 
\textit{matrix-valued  generalized Liu--Lu indicator} (see \citealp{NaszodiMendonca2021}).  
Its  $(i,j)$-th  element  is  
\begin{equation}\label{LiuLugengen}
\text{LL}^{\text{gen}}_{i,j} (Z_t)= 
\text{LL}( V_i  Z_t  W^T_j )    \;,
\end{equation}
where  $Z_t$ is the $n \times  m$ matrix representing the joint distribution;  
$V_i$ is the $2 \times  n$ matrix \vspace{6mm} \\ 
$V_i = \scriptsize{ \begin{bmatrix}
	\bovermat{\textit{i}}{1    & \cdots &  1} & \bovermat{\textit{n-i}}{ 0  & \cdots & 0}  \\
	0    & \cdots  & 0 & 1  & \cdots  & 1  	
	\end{bmatrix} }$   and  
$W^T_j$ is the $m \times 2$ matrix given by the transpose of \vspace{6mm} \\
$W_j = \scriptsize{ \begin{bmatrix}
	\bovermat{\textit{j}}{1    & \cdots & 1} & \bovermat{\textit{m-j}}{ 0  & \cdots  & 0}  \\
	0    & \cdots  & 0 & 1  & \cdots  & 1  	
	\end{bmatrix} }$ with   $ i \in \{1, \ldots, n-1 \} $, and  $j \in \{1, \ldots, m-1 \}$.  
This is how the LL-indicator is generalized for ordered, categorical, multinomial, one-dimensional assorted trait variables. 


\textbf{The NM-method}

Next, let us see how the (generalized) LL-indicator is used by the \textit{NM-method for constructing counterfactual tables}. 
We denote the NM-transformed contingency table by \\
$\text{NM}(Z_{t_p},Z_{t_a})=Z^*_{t_p, t_a}$,  
where the preferences are measured at time $t_p$, while availability is measured at time $t_a$.
Unlike  $Z_{t_p}$ and $Z_{t_a}$,  $Z^*_{t_p, t_a}$  cannot be observed. 

In the example presented in Section \ref{sec:cc},  $t_p$ corresponds to the year when the old generation is observed. 
Moreover,   $Z_{t_p}$ corresponds to table $K$ representing the joint educational distribution of couples in the old generation.  
Also,  $t_a$ corresponds to the year when the educational distribution of marriageable men and women in the young generation is observed. 
Finally,  $Z^*_{t_p, t_a}$ corresponds to table $K^{\text{yg}}$ representing the counterfactual joint educational distribution of couples in the young generation.

The counterfactual table $Z^*_{t_p, t_a}$ should meet the following two conditions:   
$\text{LL}^{\text{gen}}(Z^*_{t_p, t_a})= \text{LL}^{\text{gen}}(Z_{t_p})$ in order to make the aggregate preferences the same under the counterfactual as at time $t_p$.   
The condition on availability is given by a pair of restrictions of    
$Z^*_{t_p, t_a}  e^T_{m}=  Z_{t_a}  e^T_{m}$ and  $ e_{n} Z^*_{t_p, t_a}= e_{n} Z_{t_a}$, where $e_{m}$ and  $e_{n}$ are all-ones row vectors of size $m$ and $n$, respectively.    

First, we present the solution for $Z^*_{t_p, t_a}$  in the simplest case, where the 
assorted trait variable  is dichotomous, 
before we introduce the solution for the multinomial case. 
In the \textit{dichotomous case}, the counterfactual table $Z^*_{t_p, t_a}$ to be determined is a 2-by-2 table, just like the observed tables  
$Z_{t_p}=\begin{bmatrix}
N^p_{L,L}    &  N^p_{L,H} \\
N^p_{H,L}   &  N^p_{H,H}
\end{bmatrix}$ and 
$Z_{t_a}=\begin{bmatrix}
N^a_{L,L}    &  N^a_{L,H} \\
N^a_{H,L}   &  N^a_{H,H}
\end{bmatrix} $.  The solution for its cell corresponding to the number of  ${H,H}$-type couples is: 
\small
\begin{equation}\label{Solution}
N^*_{H,H}   =
\frac{\left[  N^p_{H,H} - \text{int}\left(\frac{N^p_{H,\cdot}N^p_{\cdot,H}} {N^p}\right)\right]  \left[{\text{min}\left(N^a_{H,\cdot}, N^a_{\cdot,H} \right)- \text{int}\left(\frac{N^a_{H,\cdot}N^a_{\cdot,H}} {N^a} \right) }\right]     }{\text{min}\left(N^p_{H,\cdot}, N^p_{\cdot,H} \right)- \text{int}\left(\frac{N^p_{H,\cdot}N^p_{\cdot,H}} {N^p} \right) }     
+\text{int}\left(\frac{N^a_{H,\cdot}N^a_{\cdot,H}} {N^a}\right)   \;, 
\end{equation}
\normalsize 
where $N^p_{H,H}$ is the number of ${H,H}$-type 
couples observed at time $t_p$. Similarly,  
$N^p_{H,\cdot}$ (the number of couples, where the husband is $H$-type),   $N^p_{\cdot,H}$ (the number of couples, where the wife is $H$-type), and  ${N^p}$ (the total number of couples) are also observed at time $t_p$; \footnote{For the derivation of Eq. (\ref{Solution}), see  \cite{NaszodiMendonca2021}.}    
whereas $N^a_{H,\cdot}$, $N^a_{\cdot,H}$, and ${N^a}$ are observed at time  $t_a$. 
So, Equation (\ref{Solution}) expresses $N^*_{H,H}$ as a function of variables with known values.  
Regarding the values of all the other three cells of $Z^*_{t_p, t_a}$, those can be calculated from $N^*_{H,H}$  by using the condition on the row totals and column totals of $Z^*_{t_p, t_a}$.

Next, let us see how the  NM-method works in the \textit{multinomial} case, where the counterfactual table  $\text{NM}(Z_{t_p},Z_{t_a})=Z^*_{t_p, t_a}$, as well as $Z_{t_p}$ and $Z_{t_a}$ are of size $n \times m$.   
It is worth to note that $\text{NM}(Z_{t_p},Z_{t_a})$ depends on the row totals and column totals of $Z_{t_a}$, but not on  $Z_{t_a}$ itself.  
So, instead of thinking of the NM-method as a function mapping $\mathbb{N}^{n \times m} \times \mathbb{N}^{n \times m} \mapsto \mathbb{R}^{n \times m}$, we should rather think of it as a function mapping $\mathbb{N}^{n \times m} \times \mathbb{N}^{n} \times \mathbb{N}^{m} \mapsto \mathbb{R}^{n \times m}$. Accordingly, we will use the following alternative notation:   
$\text{NM}(Z_{t_p}, Z_{t_a}  e^T_{m}, e_{n} Z_{t_a})$  as well. 

With this new notation,    
the problem for the \textit{multinomial, one-dimensional assortative trait} can be formalized as follows.  
Our goal is to determine the transformed contingency table $Z^*_{t_p, t_a}$ of size $n \times m$ under the restrictions 
given by the target row totals and the target column totals observed at time ${t_a}$:  $R_{t_a}:= Z_{t_a} e^T_{m}= Z^*_{t_p, t_a} e^T_{m}$, and  $C_{t_a}:= e_{n} Z_{t_a}=e_{n} Z^*_{t_p, t_a}$.   
The additional restriction is  $\text{LL}^{\text{gen}}(Z^*_{t_p, t_a})=\text{LL}^{\text{gen}}(Z_{t_p})$.

By using Eq.(\ref{LiuLugengen}), we can rewrite the problem  as follows.   
We look  for $Z^*_{t_p, t_a}$, where  
$V_i  R_{t_a} = V_i  Z^*_{t_p, t_a} e^T_{m} $, and  $C_{t_a}  W^T_j = e_{n} Z^*_{t_p, t_a} W^T_j $;  and 
$\text{LL}(V_i  Z_{t_p}  W^T_j)= \text{LL}(V_i  Z^*_{t_p, t_a}  W^T_j)$ 
for all   $i \in \{1,..., n-1\} $ and $j \in \{1,..., m-1\} $. The matrices  
$V_k$ and $W_p$ are defined the same as under Eq.(\ref{LiuLugengen}).   
For each $(i,j)$-pairs, these equations define a problem of the 2-by-2 form.       
Each problem  can be solved separately by applying Eq.(\ref{Solution}).      
The solutions determine $(n-1) \times (m-1)$ entries of the $Z^*_{t_p, t_a}$ table. 
The remaining $m+n-1$ elements of the $Z^*_{t_p, t_a}$ table can be determined with the help of the target row totals and target column totals. \\

\textbf{A numerical example}

We illustrate the application of the NM and the IPF with a numerical example. 
It shows that the transformed tables of these methods can be different.    

In our numerical example, both the husbands' and the wives' assorted trait is dichotomous taking the values low (L) or high (H).
$Z^{\text{num}}_{t_p}= {\scriptsize \begin{bmatrix}
	500   & 500  \\
	100   & 900
	\end{bmatrix} }$ 
and 
$Z^{\text{num}}_{t_a}= {\scriptsize \begin{bmatrix}
	500   & 700 \\
	100   & 700
	\end{bmatrix} }$. 

As a first step of the IPF algorithm, we factor the rows of the seed table $Z^{\text{num}}_{t_p}$ in order to match the row totals of $Z^{\text{num}}_{t_a}$.  
The table obtained after  the first step is $Q^{'}= {\scriptsize \begin{bmatrix}
	600   & 600 \\
	\;80   & 720
	\end{bmatrix} }$. 
As a second step, we factor the columns of  $Q^{'}$ in order to match the column totals of $Z^{\text{num}}_{t_a}$.  
We get  $Q^{''}= {\scriptsize \begin{bmatrix}
	529.41   & 636.36 \\
	70.59   & 763.64
	\end{bmatrix} }$. After 4  iterations, the (rounded) IPF-transformed table is  
$Z^{\text{IPF,}{\text{num}}}_{t_p, t_a}= {\scriptsize \begin{bmatrix}
	534	& 665 \\
	\;66	& 735
	\end{bmatrix} }$.


By contrast, the NM-transformed table is     
$Z^{\text{NM,}{\text{num}}}_{t_p, t_a}=  {\scriptsize \begin{bmatrix}
	520   & 680 \\
	\;80   & 720
	\end{bmatrix} }$.  

To check this result, let us first calculate  the scalar valued $\text{LL} (Z^{\text{num}}_{t_p})$ and  $\text{LL} (Z^{\text{NM,}{\text{num}}}_{t_p, t_a})$ by using Equation (\ref{LiuLusimpl}). Both are 2/3 ($\text{LL}(Z^{\text{num}}_{t_p})=\frac{\;\;900 - 700}{1,000-700}=2/3$ and  $\text{LL}(Z^{\text{NM,}{\text{num}}}_{t_p, t_a})=\frac{720 - 560}{800-560}=2/3$). In addition, the 
row totals and the column totals of $Z^{\text{NM,}{\text{num}}}_{t_p, t_a}$ are the same as those of $Z^{\text{num}}_{t_a}$. Also, by applying Equation (\ref{Solution}), we can check that the value of the $H,H$ cell of $Z^{\text{NM,}{\text{num}}}_{t_p, t_a}$ is  
$720   = \frac{\left[ 900 - 700\right]  \left[800- 560 \right]     }{ 1,000- 700 }   + 560$. 
Thereby, it is proven that $Z^{\text{NM,}{\text{num}}}_{t_p, t_a}$ is the NM-transformed table of $Z^{\text{num}}_{t_p}$ with preset marginals given by  $Z^{\text{num}}_{t_a}$.

Apparently,  table $Z^{\text{NM,}{\text{num}}}_{t_p, t_a}$ is different from  $Z^{\text{IPF,}{\text{num}}}_{t_p, t_a}$.   
Their  difference  does not vanish if we continue iterating with the IPF after the fourth step.

\vspace{2cm}

\textbf{Decomposition scheme}

As to the decomposition scheme, we apply the additive scheme with interaction effects proposed  by \cite{Biewen2014}.
For two factors ($A_{t_a}$ and $P_{t_p}$) and two time periods ($0$ and $1$), it is  
\begin{multline}\label{eq:Bdecom2}
f(A_1, P_1)-f(A_0, P_0)  = 
\overbrace{[f(A_1, P_0)-f(A_0, P_0)]}^{\mbox{\text{{\normalsize due to }}} \Delta \mbox{\textit{\normalsize A}}}+  
\overbrace{[f(A_0, P_1)-f(A_0, P_0)]}^{\mbox{\text{{\normalsize due to }}} \Delta \mbox{\textit{\normalsize  P}}}\\
+\underbrace{[f(A_1, P_1)-f(A_1, P_0) - f(A_0, P_1)+ f(A_0, P_0)]}_{\mbox{\text{\normalsize due to the joint effect of }} \Delta \mbox{\textit{\normalsize  A}} \mbox{\text{\normalsize { and }}}  \Delta \mbox{\textit{\normalsize  P}}}  \;,
\end{multline}
where function $f(A_{t_a}, P_{t_p})$ maps the space spanned by the two factors  into  $\mathbb{R}$. 
  
In the empirical analysis of Subsection \ref{sec:diff_emp},  $f(A_{t_a}, P_{t_p})$ determines the share of educationally heterogamous couples in a population, where  the structural  availability is the same as in  $A_{t_a}$, while the aggregate preferences are the same as in $P_{t_p}$. Function $f$ is the composition of function $h$ and $g$ as $f(A_{t_a}, P_{t_p})=h\circ g(A_{t_a}, P_{t_p})$, where $g(A_{t_a}, P_{t_p})$ constructs the counterfactual contingency table  if ${t_a} \neq{t_p}$, otherwise it is equal to $Z_{t_p}$. The  counterfactuals constructed for our empirical analysis are $g(A_{1980}, P_{1990})$, $g(A_{1990}, P_{1980})$, $g(A_{1990}, P_{2000})$, $g(A_{2000}, P_{1990})$, $g(A_{2000}, P_{2010})$ and $g(A_{2010}, P_{2000})$.  
While $h(Z)=  \sum_{i=1}^{n} \sum_{j=1 | j\neq i}^{m} Z_{i,j} / \sum_{i=1}^{n} \sum_{j=1}^{m} Z_{i,j}$ determines the ratio of the sum of the off-diagonal cells to the sum of all the cells of table $Z$.  
  
Depending on the choice between the NM and the IPF, the  counterfactual table $g(A_{t_a}, P_{t_p})$ for ${t_a} \neq{t_p}$  is either  $Z^{\text{NM}}_{t_p, t_a}=\text{NM}(Z_{t_p},Z_{t_a})$, or, 
$Z^{\text{IPF}}_{t_p, t_a}=\text{IPF}(Z_{t_p},Z_{t_a})$. The choice has to be made by the researcher.





\bibliography{Bib_IPF_01}

\end{document}